# Optimization and Validation of a Deep Learning CuZr Atomistic Potential: Robust Applications for Crystalline and Amorphous Phases with near-DFT Accuracy

Christopher M. Andolina, Philip Williamson, and Wissam A. Saidi*

Department of Mechanical Engineering and Materials Science, University of Pittsburgh, Pittsburgh, PA 15216, USA

*To whom correspondence should be addressed: alsaidi@pitt.edu

## Abstract

We show that a deep-learning neural network potential (DP) based on density functional theory (DFT) calculations can well describe Cu-Zr materials, an example of a binary alloy system that can coexist in several ordered intermetallics and as an amorphous phase. The complex phase diagram for Cu-Zr makes it a challenging system for traditional atomistic force-fields that fail to describe well the different properties and phases. Instead, we show that a DP approach using a large database with ~300k configurations can render results generally on par with DFT. The training set includes configurations of pristine and bulk elementary metals and intermetallics in the liquid and solid phases in addition to slab and amorphous configurations. The DP model was validated by comparing bulk properties such as lattice constants, elastic constants, bulk moduli, phonon spectra, surface energies to DFT values for identical structures. Further, we contrast the DP results with values obtained using well-established two embedded atom method potentials. Overall, our DP potential provides near DFT accuracy for the different Cu-Zr phases but with a fraction of its computational cost, thus enabling accurate computations of realistic atomistic models especially for the amorphous phase.

## I. Introduction

The development of new Earth-abundant materials is paramount for sustainability, and innovation. However, the practical discovery of new materials for commercial, medical and industrial applications has many technical challenges, is laborious, and, is generally a slow process. Copper-zirconium ($Cu_xZr_y$) is one such emerging material that has a number of curious and desirable properties[1-5] such as high strength[2, 6] and shapes memory[7]. These $Cu_xZr_y$ materials can include coexisting phases such as nanoscale crystalline clusters of different stoichiometries and bulk metallic glasses (BMG), the proportion of which vary according to temperature, pressure and/or composition.[8-9] The BMG phase exhibits remarkable temperature stability[6, 10] and interesting mechanical properties for shape memory applications[7] that are not observed in the pure ordered crystalline phases of $Cu_xZr_y$ alone. Further, the BMG properties can be further tuned by doping using Ti[11] or Al[6]. While the interplay of composition structure and functional relationships of $Cu_xZr_y$ materials imparts $Cu_xZr_y$ with many unique properties, there are still considerable challenges to understand their origins that hamper the tuning of the properties of the system.

First-principles computational methods such as density functional theory (DFT) can provide important insights and high accuracy into the optimization of these aforementioned properties. However, DFT methods alone are often not adequate for solving the structure and dynamics of complex amorphous systems. Unlike the well-defined long-range order of crystalline systems, the short-to-medium range order of amorphous systems requires simulation models with a relatively large number of atoms to minimize statistical fluctuations in the system, in addition to long simulation times to minimize the quenching rate. Both requirements pose a challenge for DFT simulations. Classical atomistic simulation methods, of which molecular dynamic simulations (MD) is the most common, have been employed in materials design for several decades to provide fast computations of atomic energies and forces. However, the parameters for these potentials to describe interatomic interactions are nontrivial to optimize. To date, the most accurate classical potentials for Cu-Zr systems are based on embedded atom method[12] (EAM) particularly those developed by Sheng *et al.* [13-15] (EAM-HS) and Mendeleev *et al.* [16] (EAM-MM) by fitting to a mixed experimental and *ab-initio* input data. While these two EAMs provide a general good description of many properties of Cu-Zr systems especially for properties that are included in the training such as the Cu-Zr melting curve as is the case of EAM-MM, they have limited transferability to configurations not included in the training as we demonstrate here. Further, and most importantly, EAM potentials have a limited functional form, and thus cannot be utilized to describe complex interactions such as for covalent and ionic bonds between Cu-Zr and reactive gasses.

Machine learning (ML) and particularly deep neural network-based force-fields have the flexibility and non-linearity necessary to describe complex potential energy surfaces.[17-23] In traditional atomistic force-field methods such as the EAMs, the complex potential energy surface is approximated using privileged functional forms that are selected based on physical motivation. On the other hand, ML potentials do not employ any explicit functional form for the dependence of the energies and forces on the atomic coordinates, but rather "learn" how atoms interact from a statistical model that relies on a massive dataset typically obtained from DFT calculations.[24-28] Similar to the traditional counterparts, ML potentials also suffer from transferability errors associated with atomic environments that are not included in the training. In fact, the transferability of ML potentials could be even worse than standard potentials given the lack of any physical intuition in the model. However, owing to the statistical learning process, ML potentials have the capacity to systematically "learn" and improve the versatility of the potential for different properties and/or regions of the materials phase space. This is a challenge for standard potentials. To-date ML-based potentials have been developed for many elemental systems such as Al[29,30], Cu[31 30], Fe[32], Zr[33-34], Mo[35], and Si[22, 36]. In contrast, fewer versatile potentials for multi-element systems exist owing to the complexity of generating the DFT database as well as the optimization of the ML force-field.

In the present study, we develop a deep neural net potential (DP) for the Cu-Zr binary alloy system using the DeepMD-Kit package[37], and systematically analyze its fidelity in describing a wide range of properties and for different phases of the system. The equilibrium phase diagram of Cu-Zr is complex with at least 6 intermetallic compounds $Cu_9Zr_2$, $Cu_{51}Zr_{14}$, $Cu_8Zr_3$, $Cu_{10}Zr_7$, $CuZr$, and $CuZr_2$ resolved experimentally in addition to bulk and liquid phases.[38] Such a phase diagram makes Cu-Zr systems a challenge for traditional force-fields. Even developing a ML potential that can accurately interpolate within the training domain of the ordered phases, in addition to the amorphous Cu-Zr phase is nontrivial.[39] In our development, we follow an adaptative iterative-learning approach to systematically augment the training dataset to circumvent data extrapolation in regions of the phase space that are of interest and are not adequately sampled. We demonstrate that the developed DP can equally describe the ordered and amorphous Cu-Zr phases.

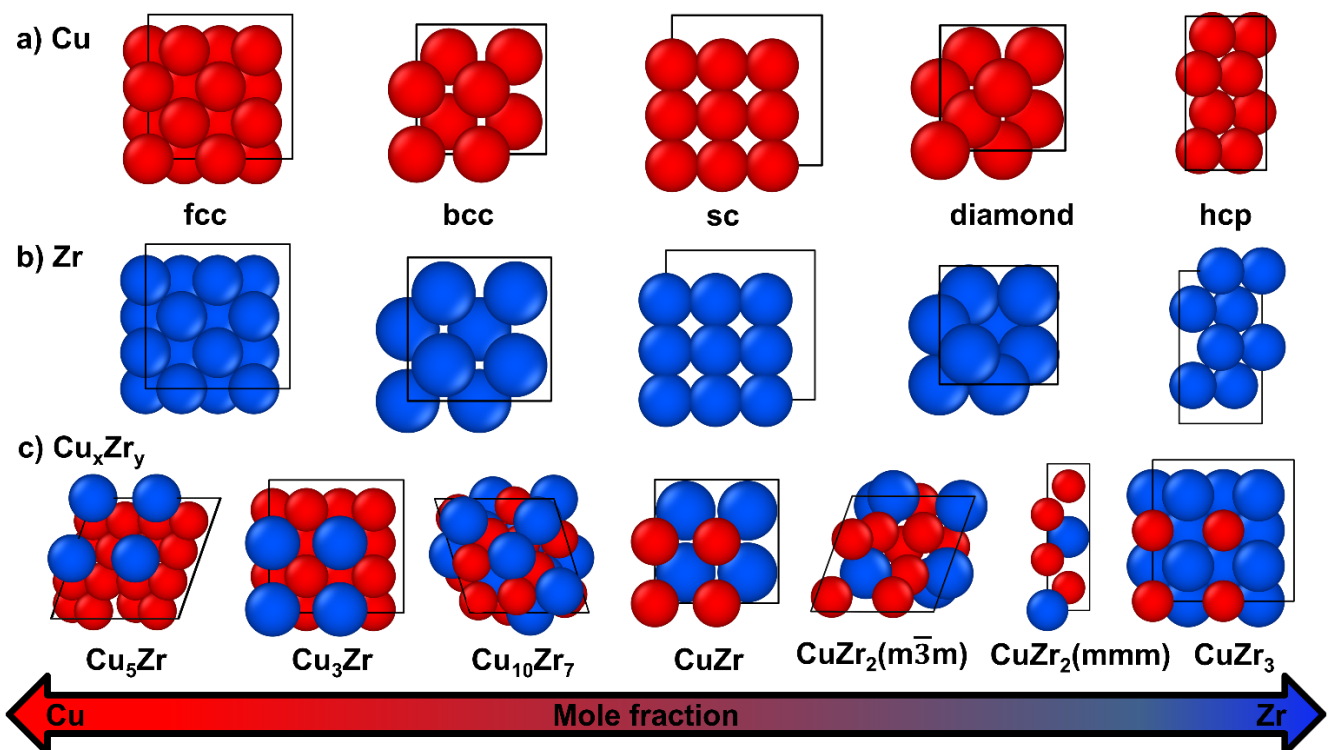

Figure 1. Atomic structures of the investigated a) Cu b) Zr and c) $Cu_xZr_y$ with the red and blue balls representing Cu and Zr, respectively.

## II.    Computation Methodology

***General Calculation Methods.*** This section describes the methods by which we model $Cu_xZr_y$ systems. See Figure 1. For the elemental metals, we investigated lattice constants, the energy of cohesion ($E_{coh}$), defects (point and planar), surface energies, and elastic constants. For the intermetallic compounds we examined their cohesive energies, lattice constants, and elastic properties. For comparison purposes to standard atomistic potentials, we select two EAM potentials from NIST EAM website[40-42] and EAM-HS *et al.[15]* We utilized the Large-scale Atomic/Molecular Massively Parallel Simulator (LAMMPS, 16 Mar 2018 version)[43] for all of our atomic calculations. The employed atomic structures in our simulations are either constructed using Atomsk[44], generated in LAMMPS or downloaded from the Material Project Database (MPDB)[45]. Figure 1 was generated using OVITO.[46]

**Bulk Crystal Lattice Constant and Atomic Energy.** Prior to investigating defects, optimized lattice constants and energy of cohesion ($E_{coh}$) are determined for the bulk crystal models. For elemental Cu and Zr systems,  we computed the cohesive energy per atom using, $E_{coh} = E_B - E_{atom}$, where $E_B$ is the bulk energy per atom and $E_{atom}$ is the energy of the corresponding isolated atom. Thus, with this definition, a negative $E_{coh}$ indicates that the system is thermodynamically stable. For the intermetallics, we computed  $E_{coh}$ with respect to bulk Cu (fcc) and Zr (hcp).

***Elastic Constants.*** In the DFT and the atomistic calculations, the elastic constants are calculated by performing 12 distortions of the lattice and then fully relaxing the atomic coordinates of the system. The elastic constants are then computed using strain-stress relationships. The bulk moduli (e.g. bulk, shear, Young moduli, and Poisson's ratios) are computed using the crystal lattice specific equations.[46-48]  For completeness, these relationships are summarized in the supporting document.

***Phonon Calculations.*** The phonon spectra are computed using a finite displacement approach utilizing the Phonopy[49] program with a 0.01Å displacement of the atom coordinates along $\pm x$, $\pm y$, and $\pm z$ directions from the equilibrium geometry. Because of symmetry, this results in only one displaced configuration. We used a 4x4x4 supercell employing conventional unit cell for Cu but we show the band structure unfolded to the primitive unit cell. For Zr, we used a 3x3x3 supercell employing the primitive unit cell. The free energy of the system is computed by Fourier interpolating on a dense 40x40x40 grid to yield smoother frequency dispersions.

**Point Defects.** *Vacancies.* Vacancy defects for Cu and Zr are modeled by taking a bulk crystal model and randomly removing an atom. This model is then optimized and the vacancy formation energy $E_{vac}$ is calculated using, $E_{vac} = E - NE_B$, where $E$ is the total energy of the optimized system with the defect and $N$ is the number of atoms. In the point defect calculations, we used superlattices with 2x2x2 unit cells.

*Self Interstitials.* The interstitial formation energies $E_f = E - (N+1) E_B$ are computed similarly to the vacancy case. We examined interstitial defects located at special symmetry points. For the octahedral (Oh) site of Cu, an adatom is inserted at the [½, ½, ½] position of the primitive unit cell. For a tetrahedral (Td) site, an atom is added to the [¼, ¼, ¼] position. For Zr, we examined interstitials located at Oh [1/3, 2/3, 1/4] and Td [0, 0, 3/8] positions either in the basal ($E_{f,bas}$) or prismatic ($E_{f,pris}$) planes of the crystal.

*Dumbbells.* Dumbbell defects (also known as split defects) were created by removing an atom and replacing it with a pair of atoms placed such that the midway point between them sits at the old location of the atom. Depending on orientation, atoms were placed such that they were equidistant between the nearest neighbor (or interstitial site) along the dumbbell axis and the other dumbbell atom. In Cu, three types of dumbbells were modeled with their axis parallel to the [100], [110], or [111] direction. In Zr, a dumbbell was created by aligning the pair axis with the [0001] direction for both the basal and prismatic structures. We have verified that after optimization the dumbbell defects retain their original orientation.

*Vacancy Mobility Energy.* The energy of the vacancy mobility was determined using the nudged elastic band method (NEB).[50] The vacancy path was mapped using 16 frames.

**Surface Energies.** The free surface formation energy $\gamma s$ is computed using $\gamma s = (E - NE_B)/(2A)$ where $E$ is the energy of the slab model and $A$ is the surface area perpendicular to the slab direction. The factor of 2 is included to account for the two different surfaces in the slab models. Because $\gamma s$ is computed from the energy difference between very thick slabs and bulk, any differences in the computational setup (e.g. k-grid density) for calculating the corresponding energies $E$ and $E_B$ could lead to erroneous results and even divergence of $\gamma s$ with slab thickness. This problem was nicely solved by Fiorentini and Methfessel who suggested to compute $\gamma s$ from the slab energies alone without the need to compute the bulk energy. In this case, $\gamma s$ is obtained by fitting the slab energies $E$ to the number of atoms, as seen from $E = 2 A \gamma s + NE_{B,fit}$ where $E_{B,fit}$ is the fitted bulk energy. [51] A key requirement for this approach is in using slabs where the convergence of $\gamma s$ is already approached.

**DFT Calculations.** The DFT database was generated using the Vienna Ab Initio Simulation Package (VASP) [52-53] employing the Perdew-Burke-Ernzerhof (PBE)[54] exchange-correlational functional to solve the Kohn-Sham equations within periodic boundary conditions. The electron-nucleus interactions are described using the projector augmented wave (PAW) method as implemented in VASP.[55] In the PAW representation, Cu is represented with a $p^1d^{10}$ valence configuration while as Zr is represented with $4s^24p^65s^24d^2$. Single-particle orbitals are expanded in plane waves generated within a cutoff of 400 eV. We use a dense gamma-centered k-grid with a 0.24 Å$^{-1}$ spacing between k-points. This is equivalent to 8x8x8 mesh for bulk Cu with conventional four atom fcc unit cell. Further, to aid in the k-grid convergence we use Methfessel-Paxton[56] of order 1 with a 0.15 eV smearing width. To ensure good convergence of energies and forces, we terminate the electronic self-consistent loop using a $10^{-8}$ eV energy-change tolerance.

***DP Training database.*** Because we aim to build a DP that can equally describe the crystalline and amorphous phases of $Cu_xZr_y$ alloys, we constructed a training database that includes bulk, surfaces, and amorphous phase. The total number of configurations in the database amounts to ~302k configurations to ensure that the intrinsically nonphysical form of the ML model has "learned" the relevant physics of the system. Most of the configurations ~200k were obtained for the small $Cu_xZr_y$ ordered compounds that have less than 10 atoms per unit cell after applying different distortions to the system. The total number of $Cu_xZr_y$ slab models was ~20k with (100), (111), (110), (211), and (321) orientations employing supercells with 20-80 atoms. The surface configurations for the alloys are obtained using fcc lattice with a Cu/Zr random occupancy. Additionally, we augmented the database with ~3k random $Cu_xZr_y$ configurations with 108 atoms per unit cell generated using Packmol[57]. The database was mainly populated from ab initio molecular dynamics (AIMD) trajectories within an NVT ensemble at a temperature that ranges between 100 and 1500 K. We employed a relatively large 2-4 femtosecond (fs) timestep in the AIMD simulations. Such a large time step is advantageous to decrease the correlations between the configurations along the AIMD trajectory, however, this incurs additional computational cost due to the increase in the number of steps to converge the Kohn-Sham wavefunction between consecutive time steps.

***DP Model.*** The DP was developed using the DeepPOT-SE approach[58] as implemented in DeePMD-Kit[37]. One of the important features of DeepPOT-SE is that it preserves translational, rotational, and permutation symmetries by employing a smooth and continuous embedding network, which is key for having a transferable and accurate representation. The DP model was tested on a wide variety of materials comprising molecular and extended systems. The extended systems include single- and multi-element metallic, semiconducting, and insulating materials.[58] Briefly, starting from the coordinate matrix $\mathfrak{R}$, DeepPot-SE first transforms $\mathfrak{R}$ to a local environment matrix $\mathfrak{R}^i$ for each atom $i$, and then each $\mathfrak{R}^i$ is mapped to a local atomic energy $E^i$ following the application of two networks. The first is an embedding network with a neural network defined filter functions $g^{i1}$ that maps $\mathfrak{R}^i$ to a list of symmetry preserving descriptors $D^i$, and the second $g^{i2}$ is a fitting network that maps $D^i$ to $E^i$ using modified feedforward network structures. The total energy $E = \sum E^i$ of the system is the sum of the obtained atomic energies $E^i$. This ensures the linear scaling of this method with system size. In our DP development, we used a cutoff radius of 7 Å for neighbor searching with 2 Å as the smooth cutoff. The maximum number of neighbors within the cutoff radius was set at 200. The dimension of the embedding and fitting nets is set at 25x50x100 and 240x240x240, respectively. The neural net is trained using Adam stochastic gradient descent method with a learning rate that decreases exponentially from the starting value of 0.001.

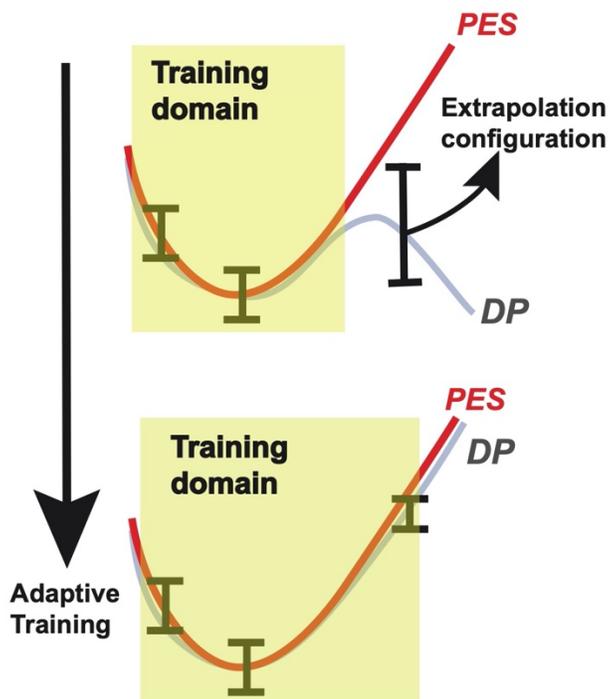

Figure 2. A schematic showing adaptive training approach for a 1-dimensional potential energy surface (PES). The red line shows the exact PES while the grey one is the fitted DP one. Extrapolative configuration can be identified from the spread of the predicted values from the DP ensemble, which are added to augment the training set utilized to generate next iteration DP.

***DP Fitting***. The input data is split into training and testing sets where the testing data is not used for optimizing the weights of the network but rather employed as an independent test for cross-validation. The training dataset comprises of the energies and forces from the DFT database, i.e. for each configuration with $N$ atoms, we have $3N+1$ reference values. The database was generated using an active learning approach to mitigate the extrapolation problem. Briefly, in this approach, the training database was selectively expanded by adding extrapolative configurations in a region of interest [39, 59-60]. In the present study, we choose an ensemble approach that trains several inequivalent networks using the inherently stochastic nature of the training process.[60-61] Given a large number of weights in the neural net, there is practically no hope in finding the global minimum, and thus the different DPs correspond to local minima that all are found different in practice. Any deviation in prediction from the ensemble of DPs to unseen data indicates a relative uncertainty in the predictability in that region of input space, as depicted schematically in Figure 2. In this case, this new configuration will be labeled using DFT, and a new DP ensemble will be generated with a larger domain of validity. The adaptive-learning loop will continue until convergence is met when no extrapolative configurations are found within the selection pool. Further, another level that we found useful is to ensure that the DP yields result as accurate as DFT for a large number of properties on systems of interest. In this regard, Cu-Zr systems are ideal given the different symmetries of bulk Cu and Zr in addition to a large number of intermetallics.

## III. Results and Discussion

To validate the DP, we investigated a wide range of properties for Cu and Zr such as point (single vacancy formation and interstitial energies), surface energies, elastic constants and moduli (*e.g.* bulk, shear, Young's moduli), and lattice phonon frequencies. Additionally, we investigated $Cu_xZr_y$ intermetallic structures ($Cu_5Zr$, $Cu_3Zr$, $Cu_{10}Zr_7$, $CuZr$, $CuZr_2$ m$\overline{3}$m, $CuZr_2$ I4/mmm, and $CuZr_3$) from structures deposited in MPDB.[62] We compared our DP results to DFT values either found in MPDB[45], literature or our own calculations. Ideally, the computed properties need to be compared with experimental values. However, here because our interest is in developing a DP model that is on par with the training dataset developed using DFT, and because of potential intrinsic DFT errors due to the exchange-correlation functional, we opted to compare with DFT rather than experiment. Further this choice is convenient because of the lack of experimental measurements for some of the properties. We also compared results from the two EAM potentials to values obtained with DFT as well. The two EAM potentials are trained by fitting to a database comprised of experimental and DFT results while the DP potential is fitted to DFT alone. Thus, an assessment of the potentials based on DFT somewhat favors the ML potential.

We quantify the errors of the potentials using a simple evaluation of the % difference with respect to the DFT values. These absolute value errors were summated and averaged divided by the number of properties ($N_{Total}$) to determine the model's predictive accuracy |Value – DFT|$_{av}$ for a specific class of property (*e.g* elastic constants, phonon frequencies, *etc.*).

***Lattice Constants, Energy of Cohesion, Elastic Constants.*** Accurate modeling of a material's mechanic properties is a fundamental first step in the validation of new potential and an area where EAM potentials have been demonstrated to perform well compared with DFT and experimental results. The lattice constants, energies of cohesion ($E_{coh}$), elastic constants and calculated moduli are listed in Table 1 for Cu and Table 2 for Zr (these calculated properties for $Cu_xZr_y$ alloys are discussed later).

*Table 1. Experimental, EAM, DP and DFT Mechanical Properties of bulk Cu.*

| Property | EXP | EAM-HS[15] | EAM-MM[42, 63] | DP | DFT |
|---|---|---|---|---|---|
| fcc a | 3.62[64] | 3.60 | 3.64 | 3.63 | 3.62 |
| fcc $E_{coh}$ | −3.48[65] | −3.54 | −3.28 | −3.69 | −3.69 |
| $C_{11}$ | 176[66] | 195 | 174 | 167 | 188 |
| $C_{12}$ | 125[66] | 130 | 128 | 121 | 132 |
| $C_{44}$ | 82[66] | 74 | 84 | 84 | 78 |
| Bulk Modulus ($G_H$) | 140[67] | 151 | 143 | 137 | 151 |
| Young's Modulus ($E_H$) | 128[67] | 152 | 158 | 156 | 166 |
| Shear Modulus ($G_H$) | 48[67] | 57 | 60 | 59 | 63 |
| Poisson's ratio (v) | 0.34[67] | 0.33 | 0.32 | 0.31 | 0.32 |
| **|Value – DFT|$_{av}$** | 0.092 | 0.669 | 0.050 | 0.058 | |

Lattice constants (a) are in Å, $E_{coh}$ in eV/atom and elastic constants and moduli in GPa. Values in italics were calculated from the measured literature values for comparison. $K_v$, $K_R$ and $K_H$ are the bulk modulus defined by Voigt, Reuss, and Hill. $G_v$, $G_R$ and $G_H$ are the Shear modulus defined by Voigt, Reuss, and Hill. $E_H$ is the calculated Young's Modulus defined by Hill (see SI for all equations). [46-48]

*Table 2. Experimental, EAM, DP and DFT Mechanical Properties of bulk hcp Zr.*

| Property | EXP | EAM-HS[15] | EAM-MM[42, 63] | DP | DFT |
|---|---|---|---|---|---|
| a | 3.23[68] | 3.20 | 3.21 | 3.23 | 3.24 |
| c | 5.15[68] | 5.23 | 5.24 | 5.15 | 5.17 |
| $E_{coh}$ | −6.24[65] | −6.32 | −6.47 | −7.20 | −7.20 |
| $C_{11}$ | 155[69] | 160 | 166 | 147 | 128 |
| $C_{12}$ | 67[69] | 73 | 66 | 87 | 83 |
| $C_{13}$ | 65[69] | 61 | 63 | 71 | 73 |
| $C_{33}$ | 175[69] | 189 | 187 | 145 | 163 |
| $C_{44}$ | 36[69] | 36 | 48 | 18 | 26 |
| $C_{66}$ | 44[69] | 44 | 50 | 28 | 30 |
| $K_v$ | 98 | 100 | 100 | 100 | 98 |
| $K_R$ | 97 | 100 | 100 | 79 | 84 |
| Bulk Modulus ($K_H$) | 98[70] | 100 | 100 | 89 | 91 |
| Young's Modulus ($E_H$) | 96[71] | 126 | 137 | 69 | 81 |
| $G_V$ | 43 | 52 | 59 | 26 | 30 |
| $G_R$ | 42 | 42 | 51 | 24 | 30 |
| Shear Modulus ($G_H$) | 37[70] | 47 | 55 | 25 | 30 |
| Poisson's ratio (v) | 0.33[70] | 0.30 | 0.27 | 0.37 | 0.35 |
| \|Value − DFT\|$_{av}$ | 0.230 | 0.257 | 0.360 | 0.090 | |

Lattice constants (a,c) are in Å. Otherwise, units and symbols are similar to Table 1.

A comparison of the mechanical properties of bulk Cu reveals similar accuracy compared to the DFT calculations for all three models EAM-HS, EAM-MM and DP with absolute |Value − DFT|$_{av}$ errors of 0.257, 0.360 and 0.090, respectively. Further, the average total difference of DFT values from the selected experimental values is 0.239, which indicates that DFT can describe experimental values relatively well. For the Zr mechanical properties, the two EAM atomistic force-fields reproduce the DFT with a similar accuracy of 0.257 and 0.360, which is nearly a factor of four larger than that of the DP model.

***Cu and Zr Additional Lattice Structures.*** Assessing a force-fields' ability to accurately optimize exotic lattice structures is another useful test to demonstrate the robustness of the potential.[13, 32, 72-73] As most of these structures are not readily observed experimentally, the ultimate utility of reproducing these values is merely for benchmarking purposes. These calculations can show the ability of the potential to distinguish between initially similar lattice spacing (*e.g.* Zr-fcc and Zr-diamond) while correctly predicting the correct final structures. Examination of the values in Table S1 shows that the two EAMs fail to accurately calculate the $E_{coh}$ (and lattice constant) of Zr-diamond. Table 3 shows ΔE, which is the change in $E_{coh}$ from the ground state structures (Cu-fcc and Zr-hcp) to the alternate structures. In general, DP yields ΔE values that are consistently similar to the DFT values (0.018, **Error! Reference source not found.**) compared to the two EAM potential values.

*Table 3.* Changes in Ground State Energy (Cu-fcc or Zr-hcp) Relative to Other Lattice Structures.

| Property | EXP | EAM-HS[15] | EAM-MM[42, 63] | DP | DFT |
|---|---|---|---|---|---|
| Cu $\Delta E_{fcc-hcp}$ | $-0.04$[65,64] | $-0.05$ | 0.20 | $-0.01$ | $-0.01$ |
| Cu $\Delta E_{fcc-bcc}$ | $-0.02$[65,64] | $-0.02$ | 0.24 | $-0.04$ | $-0.04$ |
| Cu $\Delta E_{fcc-diamond}$ | | 0.95 | 1.35 | 1.03 | 1.03 |
| Cu $\Delta E_{fcc-sc}$ | | 0.39 | 0.82 | $-0.45$ | $-0.43$ |
| Zr $\Delta E_{hcp-fcc}$ | $-0.52$[65, 74] | 0.09 | $-0.03$ | $-0.16$ | $-0.18$ |
| Zr $\Delta E_{hcp-bcc}$ | $-0.01$[65, 75] | $-0.05$ | $-0.18$ | $-0.06$ | $-0.08$ |
| Zr $\Delta E_{hcp-diamond}$ | | $-0.07$ | $-0.20$ | $-2.14$ | $-2.48$ |
| Zr $\Delta E_{hcp-sc}$ | | $-0.75$ | $-0.77$ | $-0.76$ | $-0.91$ |

***Phonon Calculations.*** The accuracy of the phonon calculations is a strong figure of merit to suggest the "global reliability" of a potential.[73, 76] As can be seen from Figure 3, the phonon spectra computed using DP compares favorably with that obtained with DFT. Here for consistency, we employed in the DFT phonon calculations the same computational setup as employed to compute the DFT database for DP training. Compared to the two EAM potentials reported in Figure S1, the quality of the DP phonon band structure is appreciably better. For Zr, the phonon band structure obtained using DP is of a lower quality compared to that of Cu. This behavior is likely because the employed training dataset sampled a significantly larger number of Cu environments compared to Zr, and also because hcp Zr with 2 atoms per unit cell has a more complex band structure compared to that of Cu with one atom per unit cell. As seen from Figure 3 (c) and (d), the free energies obtained from all atomic models are in very good agreement with those obtained with DFT with deviations less than 0.2 eV/unit cell at temperatures as high as 1750 K.

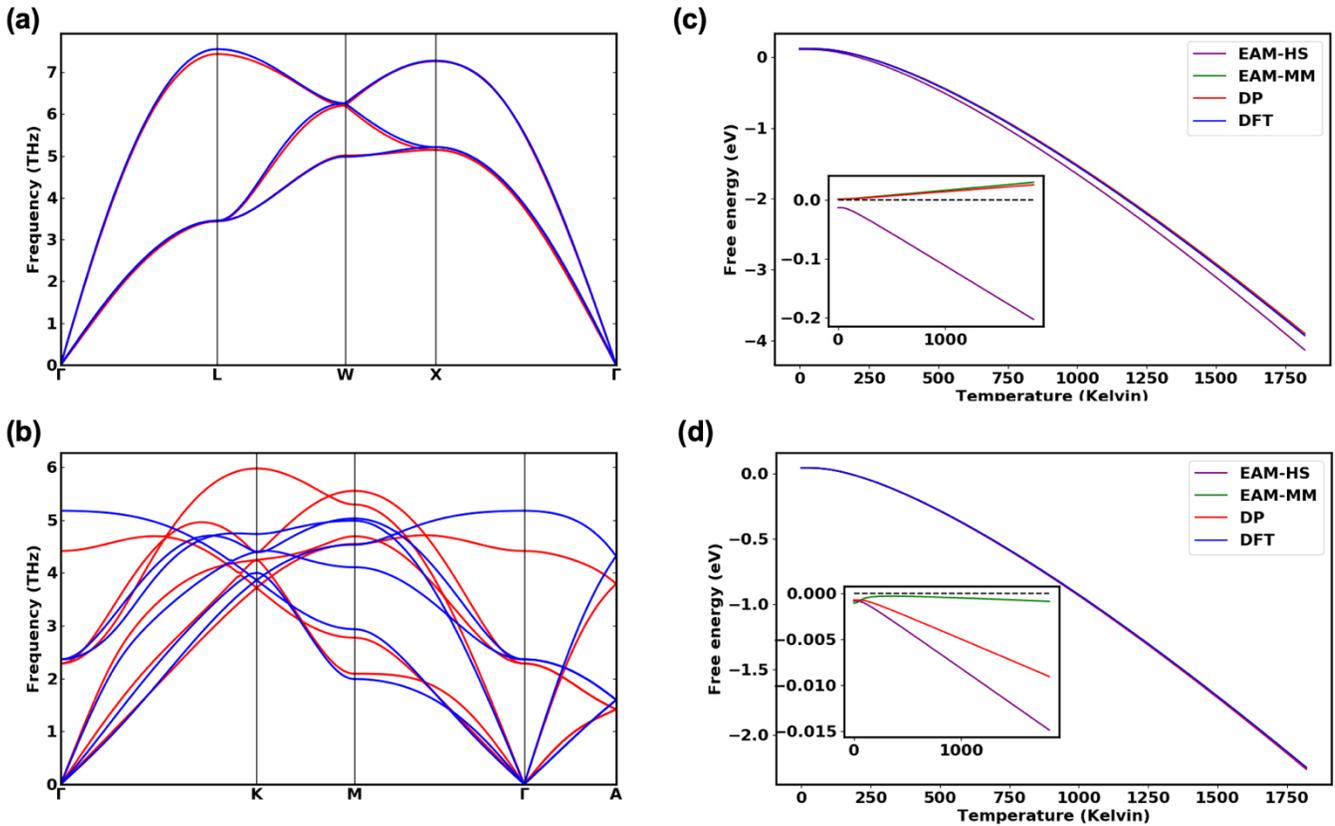

Figure 3. Comparison of phonon dispersions calculated using DFT and DP for (a) Cu-fcc, (b) Zr-hcp. Red lines indicate dispersion plot from DP and blue lines are from the DFT calculation. Corresponding comparisons for EAM-HS and EAM-MM are shown in Figure S1. (c) and (d) Comparison in the thermal energy computed from the phonon frequencies. The insets show the difference in the energy of the atomic potentials with respect to DFT values.

***Point Defects in Bulk Cu and Zr.*** Physical crystal structures are far from perfect or defect-free. Therefore, it is of great importance for theoretical models to accurately reproduce defects formation energies and atomic structures. We have investigated a selected group of point defects to evaluate the robustness of the CuZr DP potential. Namely, we determined the energies of mono-vacancies $E_{vac}$ along with their activation energies for diffusion. Also, we examined mono self- and dumbbell-interstitials. Table 4 summarizes the formation energies. Overall as seen from the table, DFT reproduces well the defects that have low formation energies but somewhat overestimates those with higher formation energies. This is consistent with other systems such as Fe[32]. The DFT diffusion energy barrier $E_{vac,mig}$ associated with the vacancy movement to an adjacent position within the lattice is in very good agreement with the experimental value. DFT predicts that Cu has a smaller diffusion barrier compared to Zr consistent with experiment (Figure 4). Also, as seen the DFT energy of the Cu[100] db is found to be the lowest energy of the db series.[77] For the Zr structures, defects located on the basal planes are generally of lower energy than those on the prismatic planes, both predicted by DFT and experimentally. Inspecting the results obtained using atomistic potentials, we see that there is excellent agreement between DP and DFT with an absolute average difference between the values of 0.066, while both EAMs perform similarly and have an order of magnitude larger error.

*Table 4. Experimental and Computed Point Defect Energies for Cu and Zr.*

| Point Defect Energy (eV) | EXP | EAM-HS[15] | EAM-MM[42, 63] | DP | DFT |
|---|---|---|---|---|---|
| Cu $E_{vac}$ | 1.20[10-11] | 1.27 | 1.05 | 1.04 | 1.16 |
| Cu $E_{vac,mig}$ | 0.65[10] | 0.81 | 1.10 | 0.79 | 0.82[78] |
| Cu $E_f$ (Oh) | 2.8[11] | 3.14 | 2.98 | 2.88 | 2.60 |
| Cu $E_f$ (Td) | 2.91[12] | 3.67 | 2.99 | 3.88 | 3.77 |
| Cu $E_f$ ([100]-db) | | 3.60 | 2.94 | 2.88 | 2.61 |
| Cu $E_f$ ([110]-db) | | 3.40 | 3.01 | 3.95 | 3.92 |
| Cu $E_f$ ([111]-db) | | 2.98 | 2.80 | 3.82 | 3.77 |
| Zr $E_{vac}$ | 1.80[25] | 1.96 | 4.53 | 2.16 | 2.05 |
| Zr $E_{vac,mig}$ | 0.58±0.04[79] | 0.63 | 1.10 | 0.69 | 0.52[80] |
| Zr $E_f$ (Oh,pris) | 3.97[26] | 3.41 | 3.14 | 3.32 | 3.41 |
| Zr $E_f$ (Oh,bas) | 2.8[81] | 2.55 | 3.14 | 3.32 | 3.41 |
| Zr $E_f$ (Td,pris) | 3.97[26] | 3.31 | 3.27 | 3.32 | 3.69 |
| Zr $E_f$ (Td,bas) | 2.8[81] | 2.59 | 3.30 | 3.50 | 3.57 |
| Zr $E_f$ ([0001]-db,pris) | | 2.59 | 3.27 | 3.63 | 3.69 |
| Zr $E_f$ ([0001]-db,bas) | | 2.78 | 3.81 | 3.50 | 3.57 |
| **|Value − DFT|$_{av}$** | 0.301 | 0.164 | 0.284 | 0.066 | |

All energies are in eV.

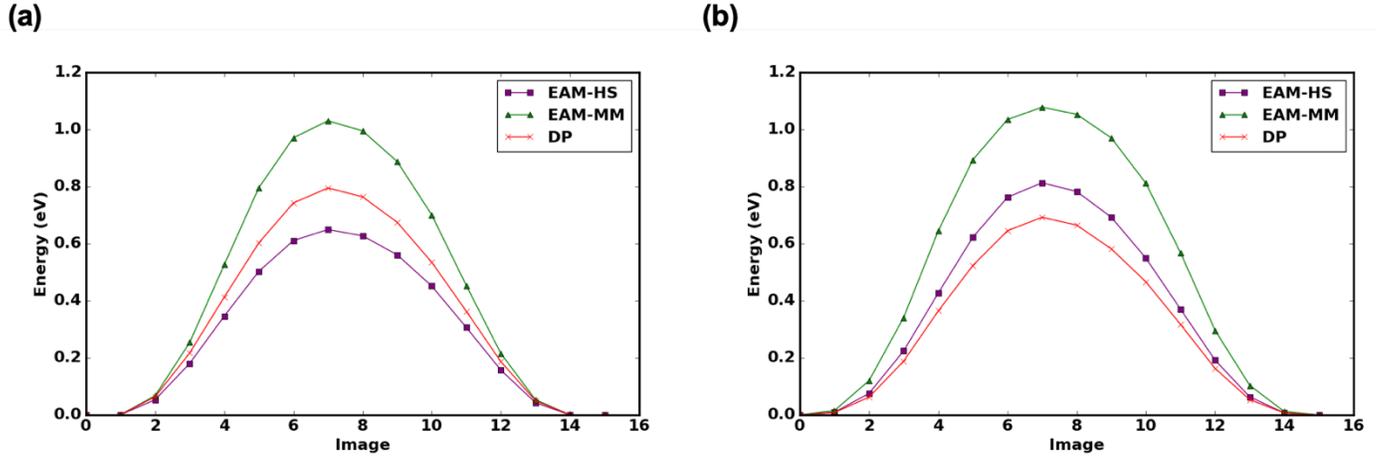

**(a)** **(b)**

*Figure 4.* Comparison of the energy barrier for vacancy migration ($E_{vac, mig}$) *via* NEB using EAM-HS, EAM-MM and DP for (a) Cu-fcc and (b) Zr-hcp.

***Free Surfaces Energies for low index Cu and Zr surfaces.*** Bulk terminated surfaces can be considered an extended defect of sorts.{Dragoni, 2018 #74} The formation energy of a solid surface is the standard method to assess surface stability. We modeled the 3 lowest Miller index surfaces for Cu (100), (110) and (111), and Zr (10-10), (11-20), and (0001). The slab models are constructed using the optimized lattice constants by orienting the surface in question perpendicular to the z-axis and doubling the z length of the simulation box to add vacuum in the non-periodic direction to mitigate spurious interactions. In our calculations, we carefully checked the convergence of surface energies. We generally attained convergence with less than 20 layers. However, we found it is important to use the optimized lattice constants to generate the surfaces, otherwise the surface energies with either the EAM or DP models are found to converge very slowly with slab thickness.

The surface energies are reported in Table 5 and we obtained similar values using either the direct or the Fiorentini-Methfessel method. As seen, for the Cu surface we are able to show the correct order of increasing surface energies 111 < 100 < 110 for all atomistic models. For Zr, also all methods show that Zr (0001) has the lowest energy, however these disagree on the two Zr (10$\overline{1}$0) than Zr (11$\overline{2}$0) surfaces. The DP values are found to agree the most with DFT compared to the two EAMs.

*Table 5. Computed Surface Energies γs in mJ/m² for Cu and Zr.*

|  | **EAM-HS**[15] | **EAM-MM**[42, 63] | **DP** | **DFT**[82] |
|---|---|---|---|---|
| Cu (100) | 1529 | 1082 | 1443 | 1470 |
| Cu (110) | 1591 | 1153 | 1496 | 1560 |
| Cu (111) | 1387 | 902 | 1202 | 1310 |
| Zr (0001) | 1263 | 1291 | 1335 | 1610 |
| Zr (10$\overline{1}$0) | 1332 | 1352 | 1422 | 1660 |
| Zr (11$\overline{2}$0) | 1494 | 1519 | 1558 | 1650 |
| **\|Value − DFT\|$_{av}$** | **0.104** | **0.217** | **0.085** | |

***Cu$_x$Zr$_y$ Intermetallic Properties.*** So far, we have demonstrated that our approach using a DP machine learning potential has either performed equally well or out performed the EAM pair style models. The ability of DF to equivalently reproduce the DFT values of common ordered Cu$_x$Zr$_y$ systems is another challenging test for the model. The optimized lattice constants and cohesive energies $E_{coh}$ of several Cu$_x$Zr$_y$ systems (Cu$_5$Zr, Cu$_3$Zr, Cu$_{10}$Zr$_7$, CuZr, CuZr$_2$ m$\overline{3}$m, CuZr$_2$ I4/mmm, and CuZr$_3$) are listed in **Error! Reference source not found.**. As seen DP continues to show excellent agreement with DFT better than either of the two EAM potentials. Overall all atomistic potentials yield results for the lattice constants and angles that are close to the DFT values. However, the two EAM potentials severely underestimate the cohesive energies although interestingly both EAMs are consistent with each other in their predictions of the values.

*Table 6. Computed lattice properties and cohesive energies for the* $Cu_xZr_y$ *compounds.*

| $Cu_xZr_y$ | Property | EXP | EAM-HS[15] | EAM-MM[42, 63] | DP | DFT[62] |
|---|---|---|---|---|---|---|
| $Cu_5Zr$ | a, α | | 4.88, 60º | 4.83, 60º | 4.88, 60º | 4.88, 60º |
| | $E_{coh}$ | | -4.05 | −3.91 | -4.96 | −4.40 |
| $Cu_3Zr$ | a | | 3.89 | 3.98 | 3.90 | 3.94 |
| | $E_{coh}$ | | −4.24 | −4.21 | −4.62 | −4.46 |
| | a, α | 9.347[83] | 9.78, 90º | 9.84, 90º | 7.85, 90º | 7.89, 90º |
| | b, β | 9.313[83] | 9.74, 90º | 9.80, 90º | 7.85, 90º | 7.89, 90º |
| $Cu_{10}Zr_7$ | c, γ | 12.675[83] | 13.19, 107º | 13.27, 107º | 9.30, 107º | 9.78, 107º |
| | $E_{coh}$ | | −5.18 | −5.22 | −5.59 | −5.59 |
| $CuZr$ | a | 3.262[84] | 3.44 | 3.41 | 3.30 | 3.27 |
| | $E_{coh}$ | | −5.58 | −5.67 | −5.57 | −5.56 |
| $CuZr_2$ ($m\overline{3}m$) | a, α | | 11.45, 60º | 11.96, 60º | 8.67, 60º | 8.69, 60º |
| | $E_{coh}$ | | -4.5 | −4.21 | -6.11 | −6.10 |
| $CuZr_2$ (I4/mmm) | a | 3.220[85] | 2.98 | 2.96 | 3.22 | 3.23 |
| | c | 11.183[85] | 10.36 | 10.26 | 11.20 | 11.20 |
| | $E_{coh}$ | | −4.5 | −4.46 | −6.17 | -6.17 |
| $CuZr_3$ | a | | 4.31 | 4.26 | 4.31 | 4.31 |
| | $E_{coh}$ | | −5.58 | −5.67 | −6.27 | −6.27 |
| **|Value − DFT|**$_{av}$ | | | 0.130 | 0.141 | 0.017 | |

Lattice constants (a,b,c) are in Å and $E_{coh}$ in eV/atom. If unit cell angles are all 90º these are not reported in the table

***Amorphous*** $Cu_xZr_y$**.** The full power of the DP model can be demonstrated by investigating how well it can describe the amorphous system. Because of the interest in comparing results to reference DFT calculations, we carried out model calculations and compared the energies of random Cu-Zr configurations. We investigated three compositions $Cu_{90}Zr_{18}$, $Cu_{60}Zr_{48}$, and $Cur_{30}Zr_{78}$ modeling Zr poor, moderate and rich conditions using a simulation cell with 108 atoms. Also, for consistency, we employed the same computational setup for the DFT calculations as used for generating the training dataset.

We examined the formation energy of the alloy $\Delta E_{alloy} = E - c_{Cu}E_B^{Cu} - c_{Zr}E_B^{Zr}$ where $E$ is the energy of the alloy pre atom, $c_{Cu}$ and $c_{Zr} = (1 - c_{Cu})$ are respectively the Cu and Zr concentrations with $E_B^{Cu}$ and $E_B^{Zr}$ the corresponding bulk energies per atom. Figure 5 compares $\Delta E_{alloy}$ obtained from DFT and the atomistic models for 25 different random realizations of the system for each of the three Cu-Zr compositions. As seen in the figure, DP accurately describes the chemical disorder in the system for the three compositions although the energy difference between some of these configurations can be as large as ~60 eV (see Figure S2). In our previous investigation of the high entropy Cantor alloy, we have also observed that the DP can accurately describe the chemical disorder.[58] Here it is important to note that these amorphous structures were not included in the training dataset. On the other hand, the results obtained using EAM-HS and EAM-MM potentials are in poor agreement with the DFT results especially for the poor and rich Zr concentrations. The results of the EAM potentials especially for the poor Zr composition shows a strong correlation between the data points suggesting that there could better agreement in the energy differences between the random configurations. While this is indeed the case, we see in Figure S2 that this is not consistent where we obtain a reasonable agreement in some configurations and errors exceeding 10 eV in others.

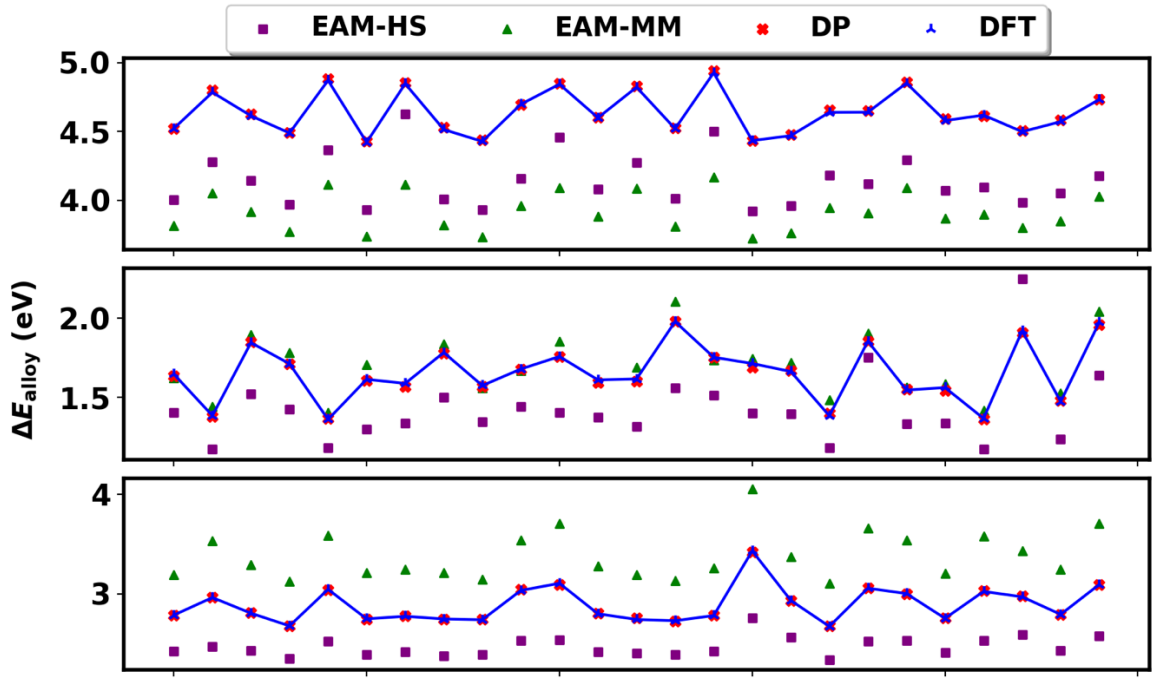

*Figure 5. Performance of different force-fields on CuZr alloy systems. The y-axis shows the difference in energy with respect to 1ˢᵗ random configuration model. The DFT data are connected with a line to aid the eye. DP is in excellent agreement with DFT.*

**Cumulative Analysis of the DP Accuracy.** Taken together our training approach of the DP potentials has indeed produced a robust EAM-style potential that approaches near DFT accuracy and is able to correctly recognize and optimize structures that were not included in this training. As with other potentials or force-fields, this model can be further appended, and the training dataset expanded to improve further the robustness and accuracy. Generally, our DP potential produces values for Cu, Zr, and $Cu_xZr_y$ that are closer to DFT values than those from established EAM potentials as summarized in *Table 7*. To be thorough we also compared calculated properties from element-specific EAM potentials (Cu and Zr) developed by Mendeleev *et al.* and Sheng *et. al.* for each monometallic property presented in the main text and still observe greater or similar accuracy from our DP potential (Tables S2 and S3).

*Table 7. Overall Comparison of the Differences from DFT for each Potential*

| Potential | Cu | Zr | $Cu_xZr_y$ | Total |
|-----------|-------|-------|------------|-------|
| EAM-HS | 0.267 | 0.201 | 0.130 | 0.220 |
| EAM-MM | 0.119 | 0.289 | 0.141 | 0.200 |
| DP | 0.039 | 0.077 | 0.017 | 0.049 |

## IV Conclusions

We developed a ML atomistic potential for Cu-Zr based on deep neural networks using a large ~300k configuration database computed using DFT. We show that the ML atomistic potential can provide a uniform accuracy over the ordered and amorphous phases of Cu-Zr that are on par with DFT results. Contrasting with two popular EAM potentials, the deep learning potential can provide an overall similar or better accuracy. The success of the DP model to describe the amorphous phases of Cu-Zr is particularly noteworthy given that this system is challenging for DFT calculations, thus enabling simulations of realistic models with near DFT-like accuracy but with a fraction of its computation cost. Further, the success of the ML potential for studying the complex Cu-Zr system motivates developing potentials for more complex interactions including ionic and covalent. We will endeavor to expand this work to other systems to provide accurate and low-cost computational tools to the material science and computational communities with these readily appended potentials.

The training database and the potential are freely available at saidigroup.pitt.edu or by contacting the corresponding author.

**ACKNOWLEDGMENTS.** We are grateful for computing time provided in part by the Pittsburgh Center for Research Computing (CRC) resources at the University of Pittsburgh, Extreme Science and Engineering Discovery Environment (XSEDE), which is supported by the National Science Foundation (NSF OCI-1053575), and Argonne Leadership Computing Facility, which is a DOE Office of Science User Facility supported under Contract DE-AC02-06CH11357.

**Supporting Information Statement.** Comparisons to atom-specific (Cu or Zr) monometallic EAMs, more details regarding the calculation of non-typical lattice configurations are listed; equations for elastic constant calculations; phonon band structures for Cu and Zr obtained using EAM-HS and EAM-MM; Convergence of surface energies of Cu and Zr with slab thickness.

## References
[1] K. Song et al., Intermetallics **67** (2015) 177.
[2] Z. Ning et al., Materials & Design **90** (2016) 145.
[3] S. Pauly et al., Journal of Materials Science & Technology **30** (2014) 584.
[4] K. Song et al., AIP Advances **3** (2013) 012116.
[5] T. H. Chen, and C. K. Tsai, Materials (Basel) **8** (2015) 1831.
[6] L. Huo et al., Journal of Non-Crystalline Solids **357** (2011) 3088.
[7] F. Qiu et al., Materials (Basel) **10** (2017) 284.
[8] H. Okamoto, Journal of phase equilibria and diffusion **29** (2008) 204.
[9] K. Zeng, M. Hämäläinen, and H. Lukas, Journal of phase equilibria **15** (1994) 577.
[10] D. Wu et al., Journal of Alloys and Compounds **664** (2016) 99.
[11] D. V. Louzguine, H. Kato, and A. Inoue, Applied physics letters **84** (2004) 1088.
[12] M. S. Daw, and M. I. Baskes, Physical Review B **29** (1984) 6443.
[13] H. W. Sheng et al., Physical Review B **83** (2011) 134118.
[14] H. W. g. w. p. Sheng, Accessed 9/1/19).
[15] Y. Q. Cheng, E. Ma, and H. W. Sheng, Phys. Rev. Lett. **102** (2009) 245501.
[16] M. I. Mendelev et al., Philosophical Magazine **89** (2009) 967.
[17] S. Lorenz, A. Groß, and M. Scheffler, Chemical Physics Letters **395** (2004) 210.
[18] B. G. Sumpter, and D. W. Noid, Chemical Physics Letters **192** (1992) 455.

[19] S. Manzhos, R. Dawes, and T. Carrington, International Journal of Quantum Chemistry **115** (2015) 1012.

[20] S. Manzhos, and T. Carrington, Jr., J. Chem. Phys. **125** (2006) 084109.

[21] A. P. Bartok *et al.*, Phys. Rev. Lett. **104** (2010) 136403.

[22] J. Behler, and M. Parrinello, Phys. Rev. Lett. **98** (2007) 146401.

[23] S. Chmiela *et al.*, Nat Commun **9** (2018) 3887.

[24] F. V. Prudente, P. H. Acioli, and J. J. S. Neto, J. Chem. Phys. **109** (1998) 8801.

[25] J. Ludwig, and D. G. Vlachos, J. Chem. Phys. **127** (2007) 154716.

[26] M. Lilichenko, and A. M. Kelley, J. Chem. Phys. **114** (2001) 7094.

[27] T. M. Rocha Filho *et al.*, International Journal of Quantum Chemistry **95** (2003) 281.

[28] T. Liu, B. Fu, and D. H. Zhang, Science China Chemistry **57** (2013) 147.

[29] G. P. P. Pun *et al.*, Nat Commun **10** (2019) 2339.

[30] T. D. Huan *et al.*, Journal of Physical Chemistry C **123** (2019) 20715.

[31] N. Artrith, and J. Behler, Physical Review B **85** (2012) 045439.

[32] D. Dragoni *et al.*, Physical Review Materials **2** (2018) 013808.

[33] X. Qian, and R. Yang, Physical Review B **98** (2018) 224108.

[34] H. Zong *et al.*, npj Computational Materials **4** (2018) 48.

[35] C. Chen *et al.*, Physical Review Materials **1** (2017) 043603.

[36] A. P. Bartók *et al.*, Physical Review X **8** (2018) 041048.

[37] G. DeePMD-Kit, 1/1/2019).

[38] D. Arias, and J. P. Abriata, Journal of Phase Equilibria **11** (1990) 452.

[39] J. Behler, J Phys Condens Matter **26** (2014) 183001.

[40] C. A. Becker *et al.*, Current Opinion in Solid State and Materials Science **17** (2013) 277.

[41] L. M. Hale, Z. T. Trautt, and C. A. Becker, Modelling and Simulation in Materials Science and Engineering **26** (2018) 055003.

[42] N. E. P. Database, Accessed 10/1/2019).

[43] S. Plimpton, Journal of Computational Physics **117** (1995) 1.

[44] P. Hirel, Comput. Phys. Comm. **197** (2015) 212.

[45] A. Jain *et al.*, APL Materials **1** (2013) 011002.

[46] A. Stukowski, Simul. Mater. Sci. Eng. **18** (2010) 015012.

[47] R. Hill, Proceedings of the Physical Society. Section A **65** (1952) 349.

[48] A. Reuss, ZAMM - Zeitschrift für Angewandte Mathematik und Mechanik **9** (1929) 49.

[49] W. Voigt, *Lehrbuch der kristallphysik* (Teubner Leipzig, 1928), Vol. 962,

[50] A. Togo, and I. Tanaka, Scripta Materialia **108** (2015) 1.

[51] G. Henkelman, and H. Jónsson, J. Chem. Phys. **113** (2000) 9978.

[52] V. Fiorentini, and M. Methfessel, Journal of Physics: Condensed Matter **8** (1996) 6525.

[53] J. F. G. Kresse, Phys. Rew. **54** (1996) 11168.

[54] M. Shishkin, M. Marsman, and G. Kresse, Phys. Rev. Lett. **99** (2007) 246403.

[55] K. B. John P. Perdew, Matthias Ernzerhof, Physical Review Letters **77** (1996) 4.

[56] G. Kresse, and D. Joubert, Physical Review B **59** (1999) 1758.

[57] M. Methfessel, and A. T. Paxton, Phys. Rev. B **40** (1989) 3616.

[58] L. Martinez *et al.*, J. Comput. Chem. **30** (2009) 2157.

[59] J. H. Linfeng Zhang, Han Wang, Wissam A. Saidi, Roberto Car, Weinan E, in *Advances in Neural Information Processing Systems 31*, edited by S. B. a. H. W. a. H. L. a. K. G. a. N. C.-B. a. R. Garnett (Curran Associates, Inc., 2018), p. 4441—4451.

[60] E. V. Podryabinkin, and A. V. Shapeev, Computational Materials Science **140** (2017) 171.

[61] L. Zhang *et al.*, arXiv:1810.11890 (2018)

[62] J. Behler, J. Chem. Phys. **145** (2016) 170901.

[63] M. P. Database, Accessed 9/1/2019).

[64] M. I. Mendelev, 2019).

[65] A. T. Dinsdale, Calphad **15** (1991) 317.

66 K. Lejaeghere *et al.*, Crit. Rev. Solid State Mater. Sci. **39** (2014)

67 W. C. Overton, and J. Gaffney, Physical Review **98** (1955) 969.

68 A. M. James, and P. M. Lord, *Macmillan's Chemical and Physical Data* (Macmillan, London, UK, 1992),

69 D. E. Gray *American Institute of Physics Handbook* (McGraw-Hill, New York, New York, 1972), 3rd edn.,

70 E. F. C. Renken, and S.-C. E. Moduli, Phys. Rev. **135** (1964) A482.

71 E. A. Brandes, and G. Brook, *Smithells metals reference book* (Elsevier, 2013),

72 S. Yamanaka *et al.*, Journal of Alloys and Compounds **293** (1999) 23.

73 B. Jelinek *et al.*, Physical Review B **85** (2012) 245102.

74 Y. Mishin *et al.*, Physical Review B **63** (2001) 224106.

75 V. L. Moruzzi, J. F. Janak, and A. R. Williams, *Calculated electronic properties of metals* (Elsevier, 2013),

76 M. Yamamoto, C. T. Chan, and K. M. Ho, Phys Rev B Condens Matter **50** (1994) 7932.

77 J. Murrell, Philosophical Magazine B **73** (1996) 163.

78 Y. Mishin *et al.*, Physical Review B **63** (2001) 224106.

79 N. Lam, L. Dagens, and N. Doan, Journal of Physics F: Metal Physics **13** (1983) 2503.

80 H. Neely, Radiation Effects **3** (1970) 189.

81 G. Vérité, F. Willaime, and C. C. Fu, in *Solid State Phenomena* (Trans Tech Publ, 2007), pp. 75.

82 C. Domain, Journal of Nuclear Materials **351** (2006) 1.

83 M. de Jong *et al.*, Scientific Data **2** (2015) 150009.

84 L. Bsenko, and J. L.-C. Met, Acta Crystallogr **15** (1962) 894.

85 E. Carvalho, and I. Harris, Journal of Materials Science **15** (1980) 1224.

86 G. A. Nevitt, and E. Flook, South Med J **55** (1962) 827.

87 M. I. Mendelev, and G. J. Ackland, Philosophical Magazine Letters **87** (2007) 349.

88 M. I. Mendelev, and A. H. King, Philosophical Magazine **93** (2013) 1268.

# Supporting Information: Optimization and Validation of a Deep Learning CuZr Atomistic Potential: Robust Applications for Crystalline and Amorphous Phases with near-DFT Accuracy

Christopher M. Andolina, Philip Williamson, and Wissam A. Saidi*

Department of Mechanical Engineering and Material Science, University of Pittsburgh, Pittsburgh, PA 15216, USA

## Table of Contents

*Table S 1. Experimental, EAM and DP, DFT values other Cu and Zr lattices.*

| Properties | Experimental | CuZrEAM[1] | CuZrEAM-FS[2] | CuZr-DP | DFT[3] |
|---|---|---|---|---|---|
| Cu-hcp (Å) a | 2.551[4] | 2.54 | 2.57 | 2.57 | 2.56 |
| Cu-hcp (Å) c | 4.190[4] | 4.41 | 4.20 | 4.19 | 4.20 |
| Cu-hcp $E_0$ (eV/atom) | -3.524[4] | -3.53 | -3.28 | -3.69 | -3.68 |
| Cu-bcc (Å) | 2.82[5] | 2.86 | 2.89 | 2.89 | 2.89 |
| Cu-bcc $E_0$ (eV/atom) | -3.50[4] | -3.50 | -3.24 | -3.65 | -3.65 |
| Cu-diamond (Å) | | 5.33 | 5.52 | 5.43 | 5.34 |
| Cu-diamond $E_0$ (eV/atom) | | -2.53 | -2.13 | -2.66 | -2.66 |
| Cu-sc (Å) | | 2.38 | 2.41 | 2.40 | 2.40 |
| Cu-sc $E_0$ (eV/atom) | | -3.09 | -2.66 | -3.24 | -3.26 |
| Zr-bcc (Å) | 3.57[6] | 3.57 | 3.56 | 3.57 | 3.58 |
| Zr-bcc $E_0$ (eV/atom) | -6.25[6] | -6.29 | -6.42 | -7.14 | -7.12 |
| Zr-fcc (Å) | 4.22[7] | 4.54 | 4.54 | 4.73 | 4.45 |
| Zr-fcc $E_0$ (eV/atom) | -6.76[8] | -6.31 | -6.44 | -7.04 | -7.02 |
| Zr-diamond (Å) | | 4.54 | 4.54 | 5.77 | 6.28 |
| Zr-diamond $E_0$ (eV/atom) | | -6.31 | -6.44 | -5.06 | -4.72 |
| Zr-sc (Å) | | 2.85 | 2.84 | 3.03 | 2.94 |
| Zr-sc $E_0$ (ev/atom) | | -5.78 | -5.75 | -6.44 | -6.29 |
| **\|Value − DFT\|_av** | 0.037 | 0.072 | 0.095 | 0.018 | |

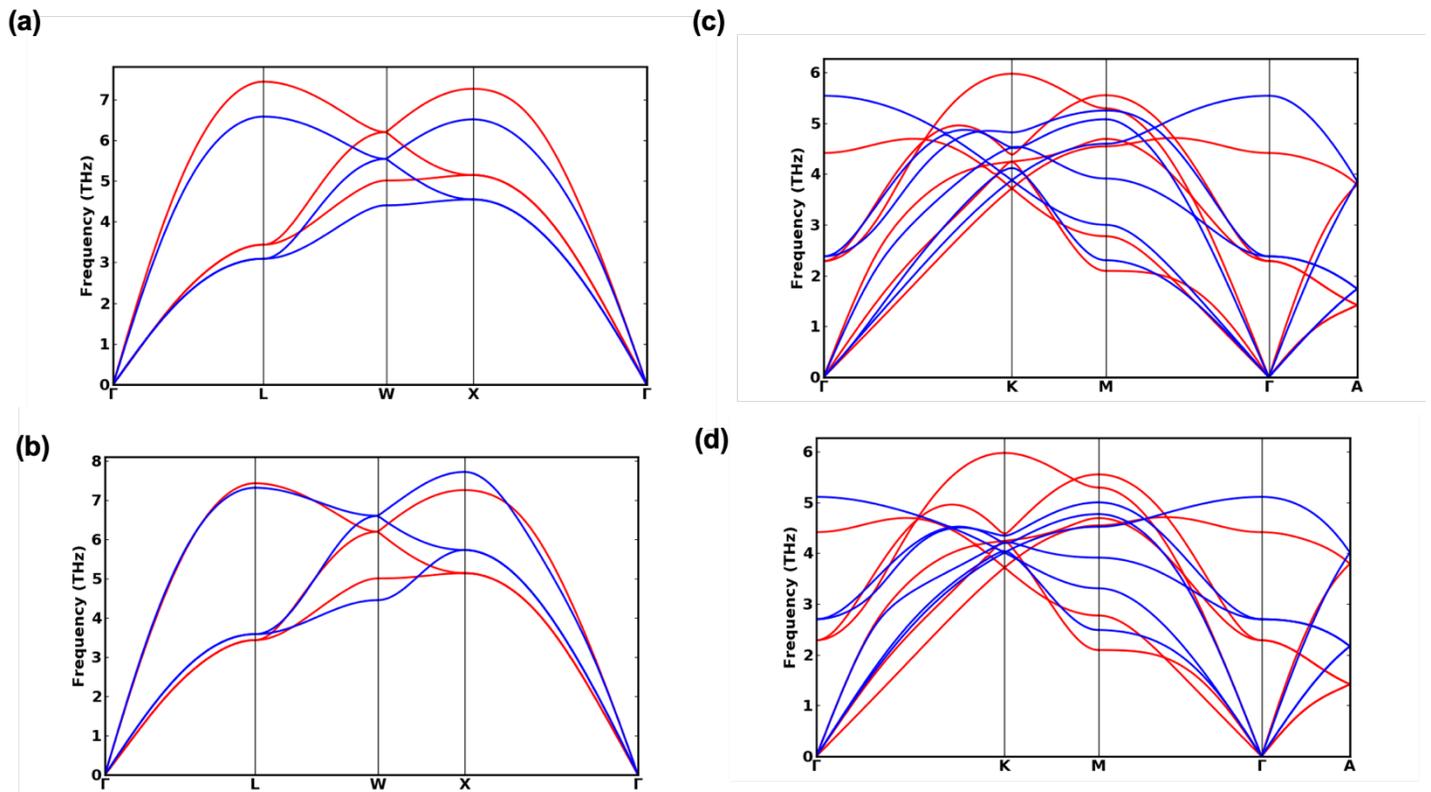

Figure S1.  Phonon band structure for Cu computed using (a) EAM-SH and (b) EAM-MM methods. The corresponding band structures for Zr are shown in (c) and (d).

*Table S 2. Calculated Properties Using EAM Cu only Potentials.*

| Property | EAM-HS-Cu | EAM-MM-Cu |
|---|---|---|
| fcc a | 3.627 | 3.627 |
| fcc $E_{coh}$ | -3.541 | -3.423 |
| $C_{11}$ | 184 | 211 |
| $C_{12}$ | 137 | 165 |
| $C_{44}$ | 83 | 86 |
| Bulk Modulus ($G_H$) | 153 | 180 |
| Young's Modulus ($E_H$) | 59 | 61 |
| Shear Modulus ($G_H$) | 158 | 164 |
| Poisson's ratio (v) | 0.33 | 0.35 |
| Cu $E_{f\ vac}$ | 1.020 | 1.140 |
| Cu $E_{f\ migration}$ | 0.795 | 0.795 |
| Cu $E_f$ (Oh) | 3.875 | 3.206 |
| Cu $E_f$ (Td) | 2.848 | 2.722 |
| Cu $E_f$ ([100]-db) | 2.889 | 2.648 |
| Cu $E_f$ ([110]-db) | 3.907 | 3.289 |
| Cu $E_f$ ([111]-db) | 3.819 | 3.166 |
| Cu $\gamma s$ (100) | 3102 | 1428 |
| Cu $\gamma s$ (110) | 3924 | 1636 |
| Cu $\gamma s$ (111) | 7558 | 2002 |
| Cu-hcp (Å) a | 2.539 | 2.566 |
| Cu-hcp (Å) c | 4.172 | 4.217 |
| Cu-hcp $E_0$ (eV/atom) | -3.534 | -3.415 |
| Cu-bcc (Å) | 2.858 | 2.894 |
| Cu-bcc $E_0$ (eV/atom) | -3.501 | -3.381 |
| Cu-diamond (Å) | 5.3272 | 5.4080 |
| Cu-diamond $E_0$ (eV/atom) | -2.531 | -2.248 |
| Cu-sc (Å) | 2.380 | 2.384 |
| Cu-sc $E_0$ (eV/atom) | -3.093 | -2.839 |
| \|Value − DFT\|$_{av}$ | 0.0140 | 0.0064 |

Lattice constants are in Å, $E_{coh}$ in eV/atom, Elastic constants, and moduli are in GPA excluding Poisson's ratio (v, unitless), point defects are in eV and surface energies are in mJ/m$^2$.

*Table S 3. Calculated Properties Using EAM Zr only Potentials.*

| Properties | EAM-HS-Zr | EAM-MM-Zr |
|---|---|---|
| hcp a | 3.22 | 3.24 |
| hcp c | 5.13 | 5.16 |
| hcp $E_{coh}$ | -6.34 | -6.63 |
| $C_{11}$ | 164 | 155 |
| $C_{12}$ | 77 | 92 |
| $C_{13}$ | 70.5 | 101 |
| $C_{33}$ | 162 | 146 |
| $C_{44}$ | 36.0 | 43.0 |
| $C_{66}$ | 45.1 | 35.9 |
| $K_V$ | 103.0 | 116 |
| $K_R$ | 95.9 | 143 |
| Bulk Modulus ($K_H$) | 99.4 | 130 |
| Young's Modulus ($E_H$) | 109 | 96.4 |
| $G_V$ | 41.9 | 36.0 |
| $G_R$ | 41.3 | 34.0 |
| Shear Modulus ($G_H$) | 41.6 | 35.0 |
| Poisson's ratio (v) | 0.32 | 0.38 |
| Zr $E_{f\,vac}$ | 1.88 | 1.67 |
| Zr $E_{f\,migration}$ | 0.69 | 0.69 |
| Zr $E_f$ (Oh) | 2.65 | 2.77 |
| Zr $E_{fbasal}$ (Oh) | 2.65 | 2.77 |
| Zr $E_f$ (Td) | 2.75 | 3.01 |
| Zr $E_{fbasal}$ (Td) | 2.72 | 3.42 |
| Zr $E_f$ ([0001]-db) | 3.01 | 2.77 |
| Zr $E_{fbasal}$ ([0001]-db) | 2.72 | 3.42 |
| Zr $\gamma_S$ (0001) | 1318 | 1515 |
| Zr $\gamma_S$ (1010) | 1186 | 1563 |
| Zr $\gamma_S$ (1120) | 1387 | 1692 |
| Zr-bcc (Å) | 0.45 | 0.45 |
| Zr-bcc $E_0$ (eV/atom) | -6.27 | -6.53 |
| Zr-fcc (Å) | 0.57 | 0.57 |
| Zr-fcc $E_0$ (eV/atom) | -6.31 | -6.58 |
| Zr-diamond (Å) | 0.74 | 0.76 |
| Zr-diamond $E_0$ (eV/atom) | -4.82 | -5.34 |
| Zr-sc (Å) | 0.36 | 0.36 |
| Zr-sc $E_0$ (ev/atom) | -5.77 | -6.22 |
| $|Value - DFT|_{av}$ | 0.240 | 0.234 |

Lattice constants are in Å, $E_{coh}$ in eV/atom, Elastic constants, and moduli are in GPA excluding Poisson's ratio (v, unitless), point defects are in eV and surface energies are in mJ/m$^2$

**General Elastic Moduli Equations.** Below is a list of the equations used to calculate elastic moduli (*e.g* Bulk, Shear, Young's and Poisson ratio). All of these equations are from the works of Voigt, Reuss, and Hill. [9-11]

$$K_H = \frac{K_V + K_R}{2}$$

$$G_H = \frac{G_V + G_R}{2}$$

$$E_H = \frac{9K_H\, G_H}{3K_H + G_H}$$

$$v = \frac{3K_H - 2G_H}{6K_H + 2G_H}$$

**Cubic systems**

$$K_{V,G} = \frac{C_{11} + 2C_{22}}{3}$$

$$G_V = \frac{3C_{44} + C_{11} - C_{12}}{5}$$
$$G_R = \frac{5C_{44}(C_{11} + C_{12})}{4C_{44} + 3(C_{11} - C_{12})}$$

**Hexagonal systems**

$$K_V = \frac{2C_{11} + 2C_{12} + 4C_{13} + C_{33}}{9}$$

$$G_V = \frac{C_{11} + C_{12} + 2C_{33} - 4C_{13} + 12C_{44} + 12C_{66}}{30}$$

$$K_V = \frac{(C_{11} + C_{12})C_{33} - 2C_{12} + 4C_{13} + C_{33}}{C_{11} + C_{12} + 2C_{33} - 4C_{13}}$$

$$G_R = \left(\frac{5}{2}\right)\frac{\{[(C_{11} + C_{12})C_{33} - 2C_{12}](C_{55}C_{66})\}}{\{3K_V C_{55}C_{66} + [(C_{11} + C_{12})C_{33} - 2C_{12}]^2(C_{55} + C_{66})\}}$$

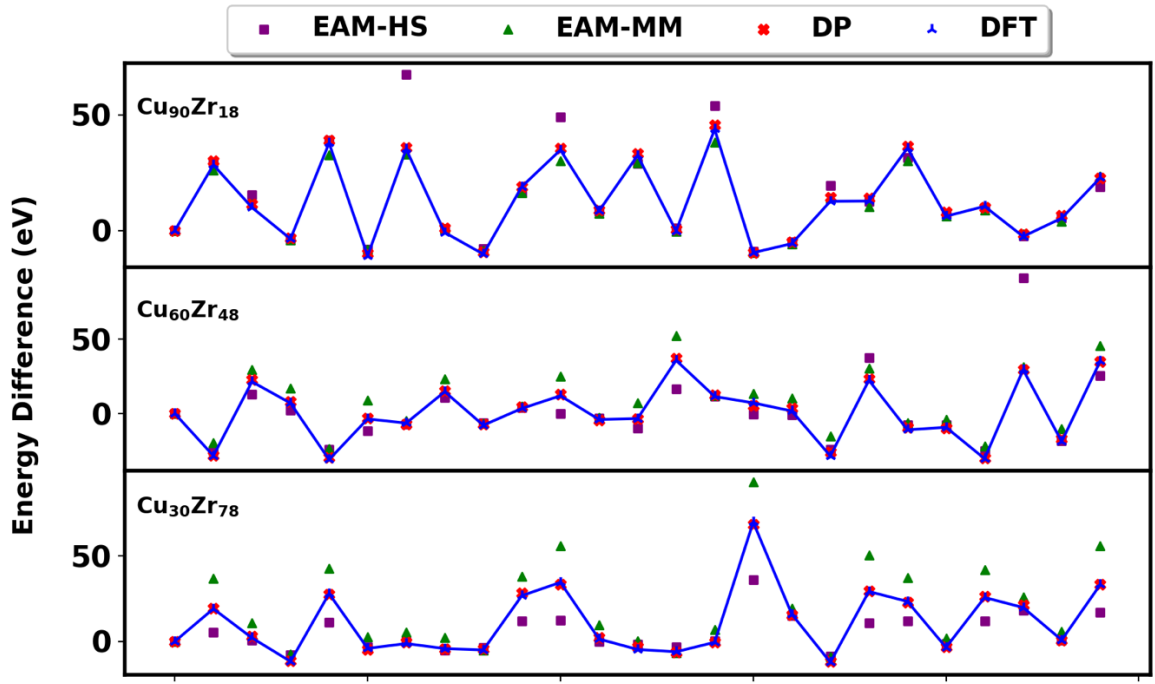

*Figure S2. Performance of different force-fields on CuZr alloy systems. The y-axis shows the difference in the total energy with respect to 1ˢᵗ random configuration model. The DFT data are connected with a line to aid the eye. DP is in excellent agreement with DFT. The two EAM potentials have absolute errors in excess of 10 eV for some configurations.*

**References.**

[1] Y. Q. Cheng, E. Ma, and H. W. Sheng, Phys Rev Lett **102** (2009) 245501.
[2]
[3] M. de Jong *et al.*, Sci Data **2** (2015) 150009.
[4] A. T. Dinsdale, Calphad **15** (1991) 317.
[5] I. A. Morrison, M. H. Kang, and E. J. Mele, Phys Rev B Condens Matter **39** (1989) 1575.
[6] F. Willaime, and C. Massobrio, Phys Rev B Condens Matter **43** (1991) 11653.
[7] M. Yamamoto, C. T. Chan, and K. M. Ho, Phys Rev B Condens Matter **50** (1994) 7932.
[8] V. L. Moruzzi, J. F. Janak, and A. R. Williams, *Calculated electronic properties of metals* (Elsevier, 2013),
[9] R. Hill, Proceedings of the Physical Society. Section A **65** (1952) 349.
[10] A. Reuss, ZAMM - Journal of Applied Mathematics and Mechanics / Zeitschrift für Angewandte Mathematik und Mechanik **9** (1929) 49.
[11] W. Voigt, *Lehrbuch der kristallphysik* (Teubner Leipzig, 1928), Vol. 962,